\documentclass[a4paper,11pt]{article}
\pdfoutput=1 

\usepackage{epsfig}
\usepackage{graphicx}                  
\usepackage{ae}
\usepackage{amsmath}
\usepackage{amssymb}
\usepackage{graphics}
\usepackage{dcolumn}
\usepackage{bm}
\usepackage{lineno}
\usepackage{caption}
\usepackage{ragged2e}
\usepackage{subcaption}

\renewcommand{\thefootnote}{\fnsymbol{footnote}}
\title{\bf Stability study of GEM chamber using radioactive source}
\date{}

\begin{document}

\maketitle
	\flushbottom
\vspace*{-1cm}
\centering

\author{\bf S.~Mandal,}
\author{\bf S.~Gope,}
\author{\bf S.~Das,}
\author{\bf S. Biswas$^*$}
\let\thefootnote\relax\footnotetext{$^*$Corresponding author. 

\hspace*{0.4cm}E-mail: saikat@jcbose.ac.in }

\vspace*{0.5cm}

	{{Department of Physical Sciences, Bose Institute, EN-80, Sector V, Kolkata-700091, India}

\vspace*{0.5cm}
\centering{\bf Abstract}
\justify

Gas Electron Multiplier (GEM) is a cutting edge Micro Pattern Gaseous detector (MPGD) technology suitable as tracking device in high rate Heavy-Ion (HI) experiments for their good spatial resolution and most importantly high rate handling capability. The performance studies including the detector efficiency, gain, energy resolution and also the stability study under high radiation are most important aspects, to be investigated before using the detector in any experiment. In this work, all of the above mentioned aspects are investigated using a $^{55}$Fe X-ray source for a single mask triple GEM chamber prototype operated with premixed Argon/CO$_2$ (Ar/CO$_2$) gas mixture in $70/30$ volume ratio. In this article, particularly the stability in efficiency using a radioactive source is discussed in detail.






\vspace*{0.25cm}





\section{Introduction} \label{intro}
GEM detectors invented by Fabio Sauli in 1997, are widely used in many High Energy Physics (HEP) experiments as tracking devices due to their high rate handling capability and good position resolution \cite{Sauli}.GEM are used as particle detectors in a variety of scientific fields, including astronomy, nuclear and particle physics, and medical diagnostics \cite{Bucciantonio}. The Compressed Baryonic Matter (CBM) experiment \cite{cbm} at the future Facility for Antiproton and Ion Research (FAIR) \cite{fair}, Darmstadt, Germany, will use the triple GEM detector as a tracking device in the Muon Chamber (MuCh) \cite{Galatyuk,Buzulutskov,Ketzer} to track di-muon pairs originating from the decay of $J/\psi$ and low-mass vector mesons ($\rho$, $\omega$, $\phi$). Triple GEM detectors will be  implemented in the first two stations of CBM-MuCh because a very high particle rate ($\sim 1.0\ \mathrm{MHz/cm^2}$ for the first station ) is expected in these regions \cite{Chatterjee_thesis}. The stability in terms of gain and energy resolution of a prototype Single Mask (SM) triple GEM detector is investigated with high-rate X-ray irradiation. Premixed Ar/CO$_2$ gas in $70/30$ volume ratio is used for this study. A strong $^{55}$Fe X-ray source is used to irradiate the chamber, and the same source is also used to monitor the spectra. A small area of the chamber ($50~\text{mm}^2$) is continuously exposed to the X-ray throughout the operation \cite{chatterjee_2023_rh,s_chatterjee_charging_up_1,s_chatterjee_charging_up_2,uniformity_1,chatterjee_2023_charge}. The effect of temperature and pressure on the gain and energy resolution is monitored. In earlier studies an interesting behaviour of the variation of the gain and energy resolution with the variation of the bias current is observed and also reported \cite{chatterjee_2023_jinst, mandal_2024}. The motivation of this study is also to understand the behaviour of the chamber under continuous high irradiation. Particularly, the stability of the chamber's efficiency (count rate with a radioactive source) and initial effects are discussed in this article.

\section{Detector configuration and experimental setup} \label{det_con}
The SM triple GEM detector prototype, consisting of $10\ \mathrm{cm} \times 10\ \mathrm{cm}$ standard GEM foils produced at CERN, is assembled in the RD51 laboratory \cite{adak}. The chamber is configured with a drift gap of $3\ \mathrm{mm}$, two transfer gaps of $2\ \mathrm{mm}$ each, and an induction gap of $2\ \mathrm{mm}$ \cite{mandal_2024}. A resistive voltage divider chain is used to power the chamber, as illustrated in Fig.~\ref{fig1}.
\begin{figure}[htb!]
	\begin{center} 
	\includegraphics[scale=0.35]{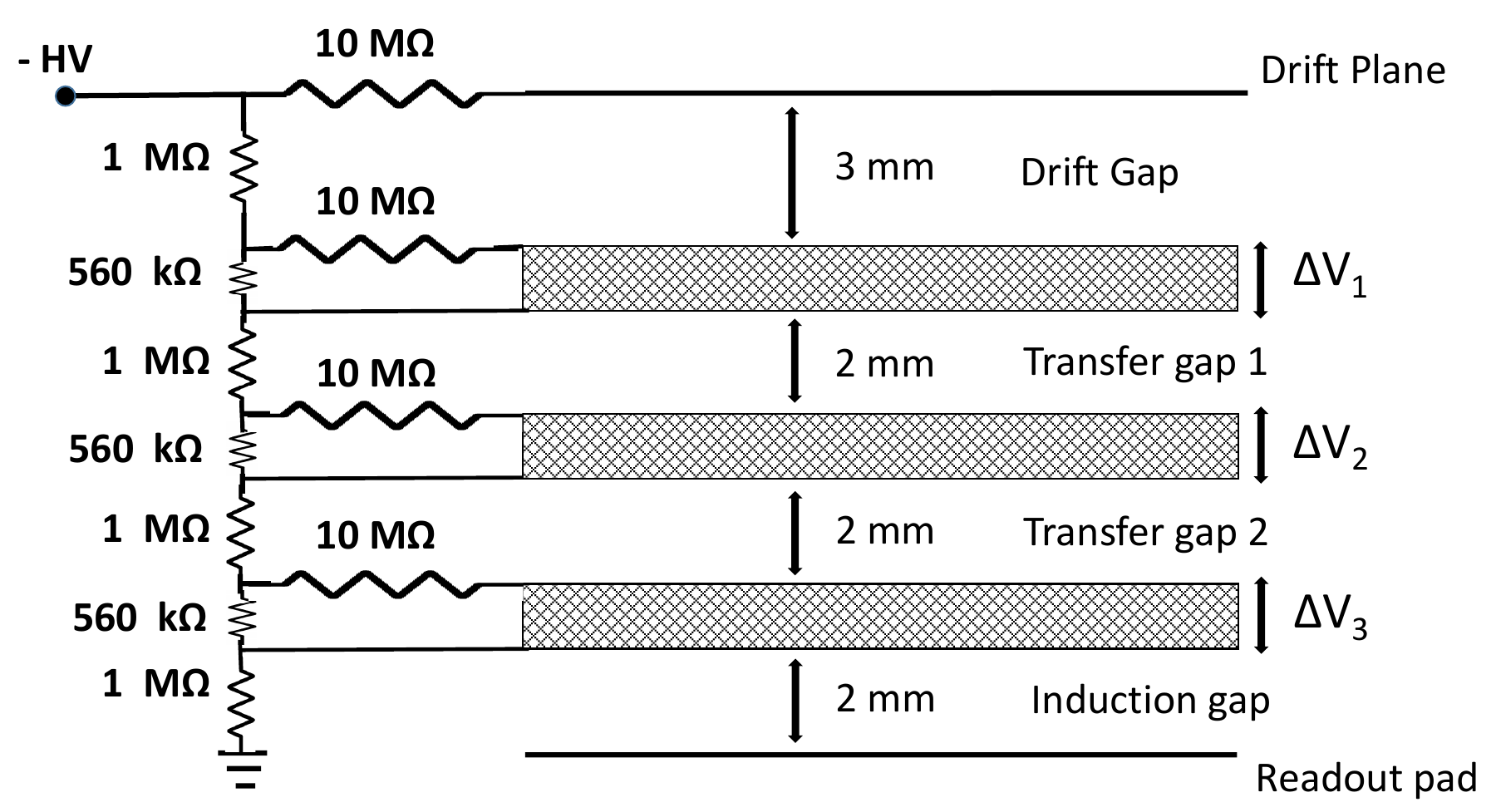}
	\caption{Schematic of the HV distribution of the SM triple GEM chamber \cite{,s_chatterjee_charging_up_1}.}\label{fig1}
	\end{center}
\end{figure}
\begin{figure}[htb!]
	\begin{center} 
	\includegraphics[scale=0.5]{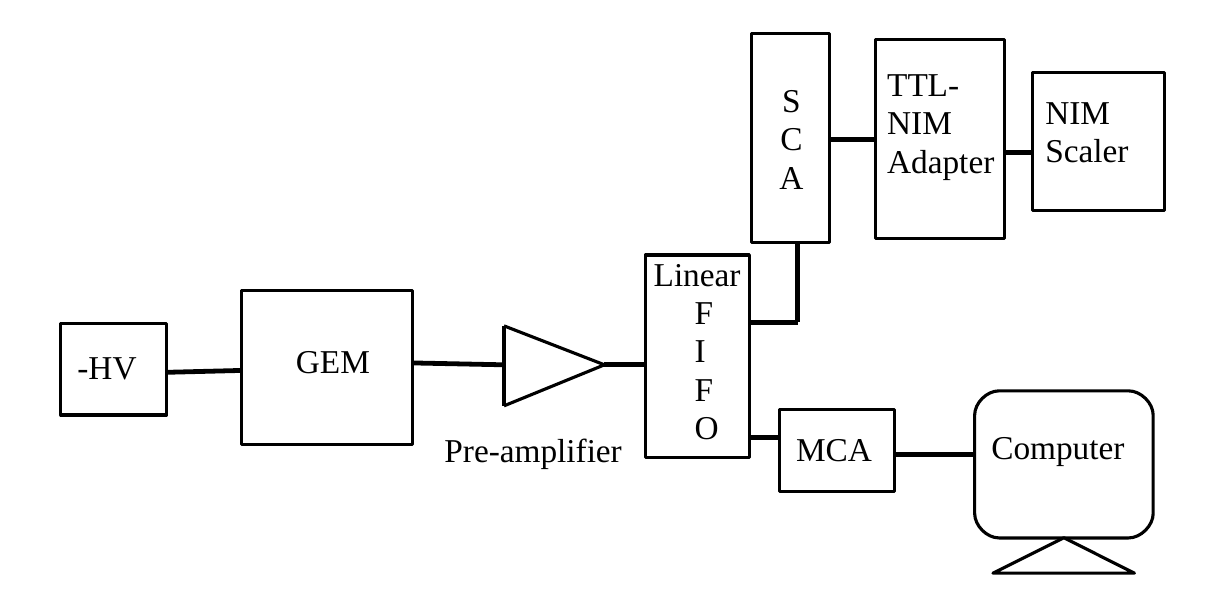}
	\caption{ Schematic representation of the electronic setup \cite{,s_chatterjee_charging_up_1}.}\label{fig2}
	\end{center}
\end{figure}
An XY printed circuit board (256 X-tracks and 256 Y-tracks) on the base plate, serves as the readout plane. Each set of 256 tracks is connected to two Panasonic 128-pin connectors. It is to be mentioned here that in this study, signals from individual readout strips are not collected separately. Instead, a sum-up board (provided by CERN) is used for each Panasonic connector. Total four sum-up boards are used for the prototype. In this study, one sum-up board is used for signal collection, and other three sum-up boards are terminated with $50~\Omega$ resistors. The signal from the sum-up board is fed into a charge-sensitive preamplifier (VV50-2) having a gain of 2 mV/fC and shaping time of 300 ns \cite{preamplifier}. The output of the preamplifier is connected to a linear Fan-in-Fan-out (FIFO) module. One analog output from the FIFO is directed to a Single Channel Analyzer (SCA), which is operated in integral mode to measure the rate of incident X-ray. The lower-level threshold to the SCA is set to 0.9 V to filter out the noise. The discriminated TTL signal from the SCA is converted to a NIM signal using a TTL-NIM adapter, and the output is counted using a scaler. Another analog output from the linear FIFO is sent to a Multi-Channel Analyzer (MCA) to record the energy spectra in a computer. The schematic of the electronic circuit is shown in Fig.~\ref{fig2}. In this study, the chamber is operated with premixed Ar/CO$_2$ gas in $70/30$ volume ratio. A constant gas flow rate of 3.5 l/h is maintained using a Vögtlin gas flow meter. A collimator made of G-10 material is used to irradiate a particular patch of the chamber with X-ray flux from $^{55}$Fe source. Ambient temperature, pressure, and relative humidity are continuously monitored using an in-house-built data logger \cite{sahu}

\section{Observables}
In this study, the gain, energy resolution, count rate, divider current, ambient temperature (t), pressure (p) and relative humidity (RH) are measured in a periodic manner. The gain and energy resolution are calculated by fitting the 5.9 keV peak of the $^{55}$Fe X-ray spectrum using the Gaussian function. The gain is defined by 

\begin{equation}
\begin{aligned}
gain  
& = \frac{output \; charge}{input \: charge} & = \frac{\frac{pulse \; hight}{2 \: mV} fC}{no.\; of \; primary \; electrons\; \times \; eC} 
\end{aligned}
\end{equation}

From the spectrum, the $pulse~height$ in Volt is found out using the MCA calibration factor. The MCA is calibrated before hand using a known signal from a function generator. The $pulse~height$ is extracted from the mean MCA channel number of the Gaussian fitted 5.9~keV peak, using the relation,

\begin{equation}
\begin{aligned}
pulse \; height \; (V) & = MCA \; Channel \; no. \times 0.001 \; + \;  0.14
\end{aligned}
\end{equation}
The details of the MCA calibration is described in Ref~\cite{Chatterjee_thesis}.

The full width at half maxima (FWHM) of the Gaussian fitted spectrum is used to define the energy resolution of the chamber using the relation,

\begin{equation}
\centering
Energy \; resolution = \frac{sigma \times 2.355}{mean}   
\end{equation}
where the $sigma$ and the $mean$ are obtained from the Gaussian fitted 5.9~keV peak of the $^{55}$Fe energy spectra.\par
While detector characterisation, one can measure the highest efficiency regime of a gas detector, measuring the count rate using a radioactive source. A radioactive source usually emits constant numbers of particles at a particular time interval. If the applied voltage of the detector is increased it can be seen the measured count rate is also increasing. After a particular applied voltage the measured count rate saturates. It means the detector reached its highest efficiency value. In this article, the measurement of the highest efficiency level and its stability with time is emphasised.

\section{Results}
The results of the stability test of the SM triple GEM detector are presented in this section. Initially, the negative high voltage (HV) is set to 4300 V. Using a collimator, $50~\text{mm}^2$ area of the prototype is exposed to an X-ray radiation and the spectra are recorded for every minute. The temperature, pressure, and relative humidity are also recorded at the same interval. The divider current is manually recorded from the HV power supply. The gain, energy resolution, bias current and T/p (T=t+273) are plotted as a function of time in Fig.~\ref{fig3}.
\begin{figure}[htb!]
	\begin{center} 
	\includegraphics[scale=0.5]{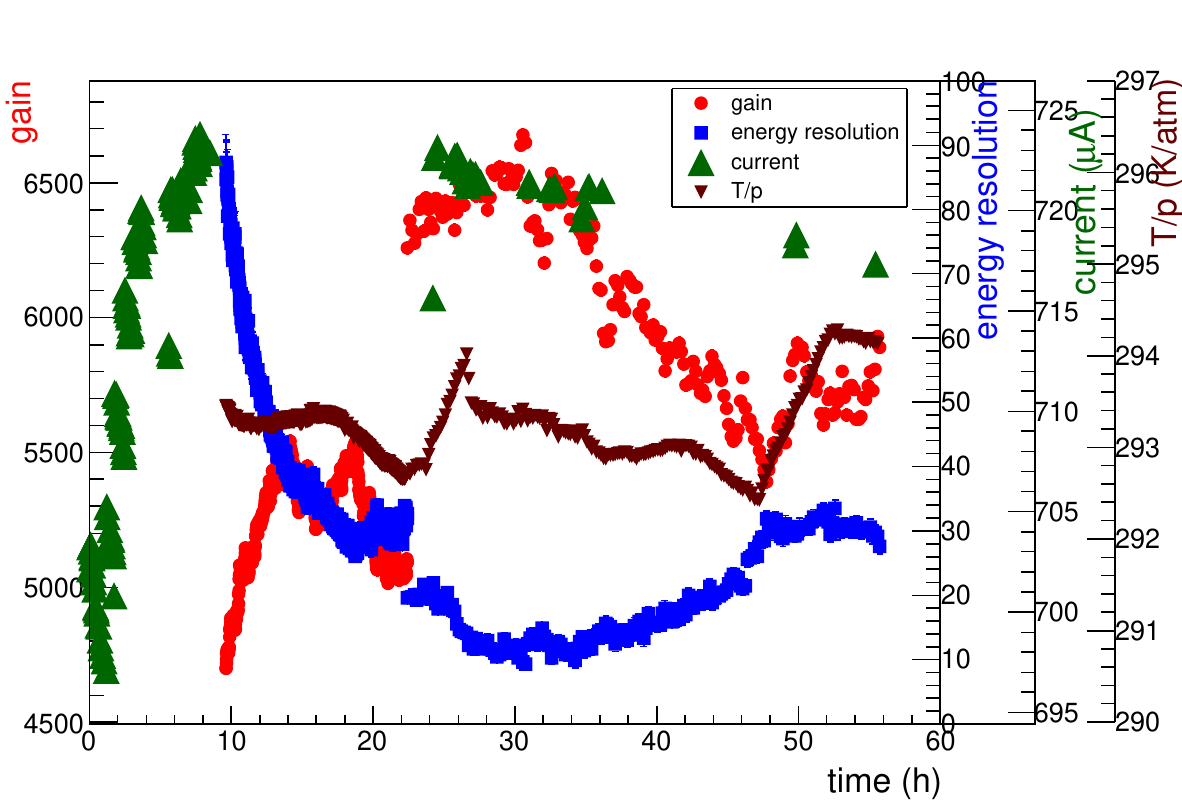}
	\caption {(Colour online )Variation of the gain, energy resolution, T/p and current as a function of time.}\label{fig3}
	\end{center}
\end{figure}
As shown in Fig.~\ref{fig3}, immediately after applying the HV, the current increases rapidly, reaches a maximum and then begins to decrease. First 10 hours at 4300 V the spectra are found to be not in a good shape and it was difficult to fit the main peak with the Gaussian function. As a result the gain and energy resolution are not possible to be calculated for those initial period. However, the count rate is measured from the beginning. After about 10 hours from the start of operation good X-ray spectra are obtained and accordingly gain and energy resolutions are calculated and the values can be seen in Fig.~\ref{fig3}. It is also clear from Fig.~\ref{fig3} that beyond
\begin{figure}[htb!]
	\begin{center}
	\includegraphics[scale=0.5]{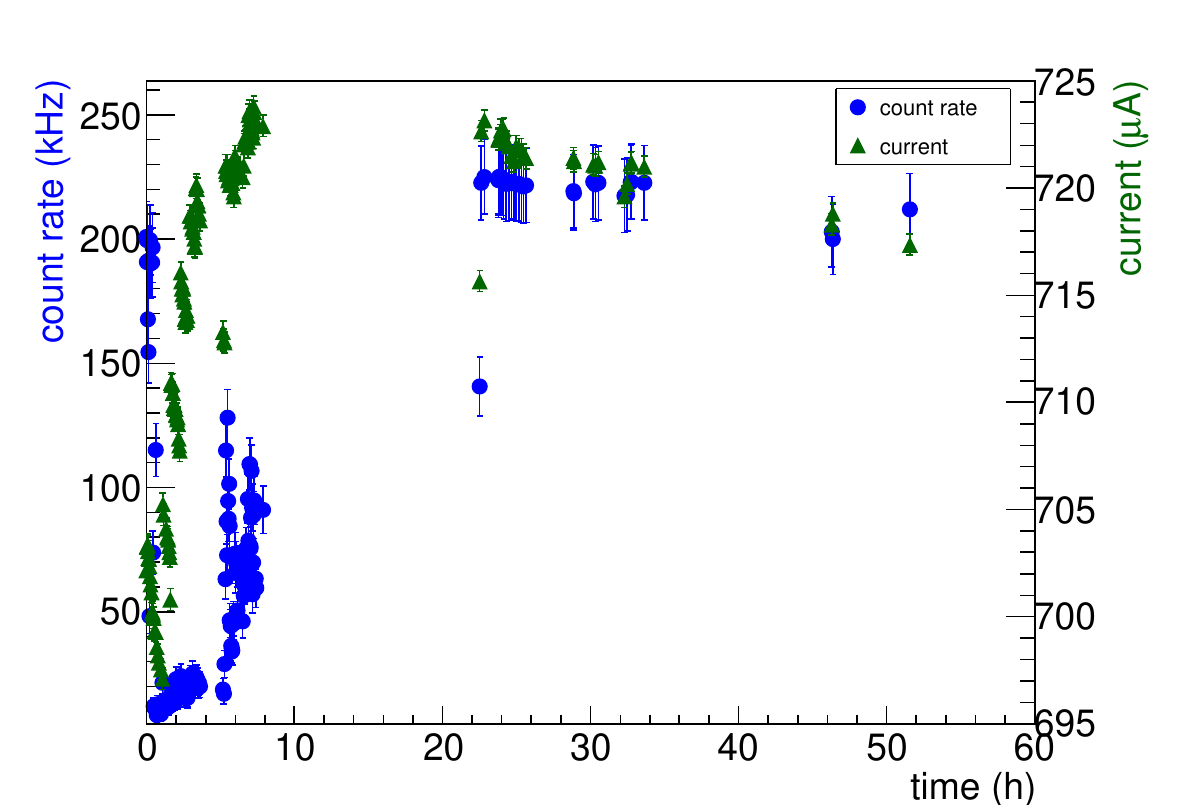}
	\caption{(Colour online) Variation of the count rate and current as a function of time.}\label{fig4}
	\end{center}
\end{figure}
10 hours the gain shows an inverse correlation with energy resolution as expected. While temperature and pressure fluctuations contribute to the gain variation, the continuous decrease in bias current is also identified as another important factor for the gain decrease. To keep the gain stable the applied voltage is increased such that the divider current can be increased, which in turn will increase the $\Delta V$. Similar study in detail is also reported in Ref. \cite{mandal_2024}. \par
\begin{figure}[htb!]
	\begin{center} 
	\includegraphics[scale=0.5]{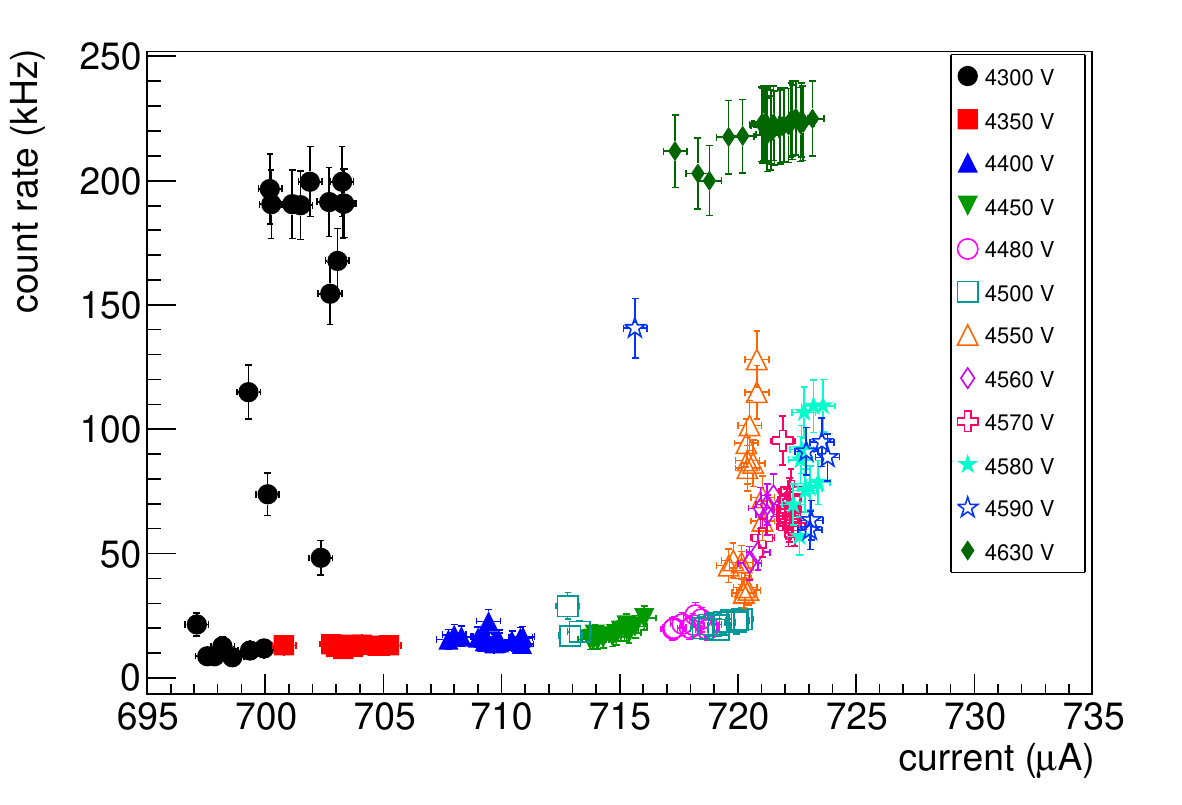}
	\caption{(Colour online) Variation of the count rate as a function of current}\label{fig5}
	\end{center}
\end{figure}
In this particular study, in addition to the gain and energy resolution, the variation in the count rate with time is also studied using a radioactive source. In Fig.~\ref{fig4} both count rate and divider current is plotted as a function of time.
\begin{figure}[htb!]
	\centering 
	\begin{subfigure}[b]{0.495\textwidth}
	\includegraphics[width=\linewidth]{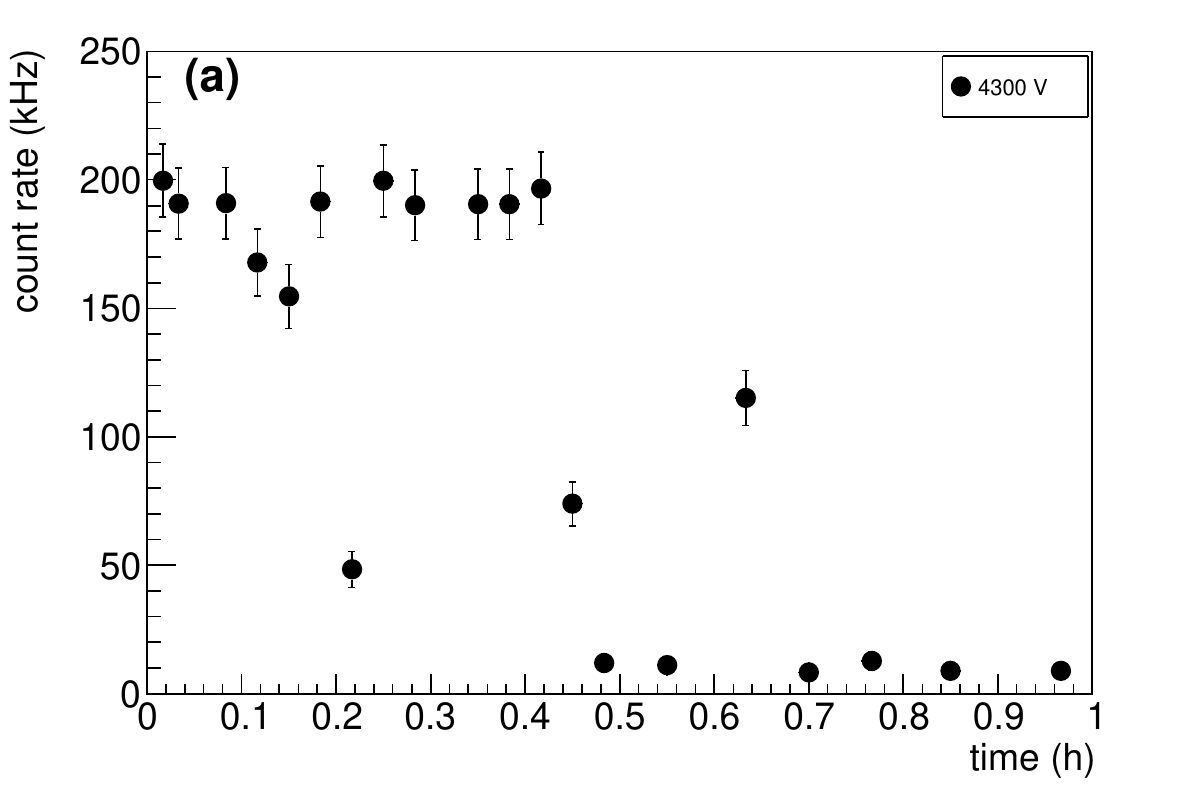}
	\label{fig:sub1}
	\end{subfigure}
	\hfill
	\begin{subfigure}[b]{0.495\textwidth}
	\includegraphics[width=\linewidth]{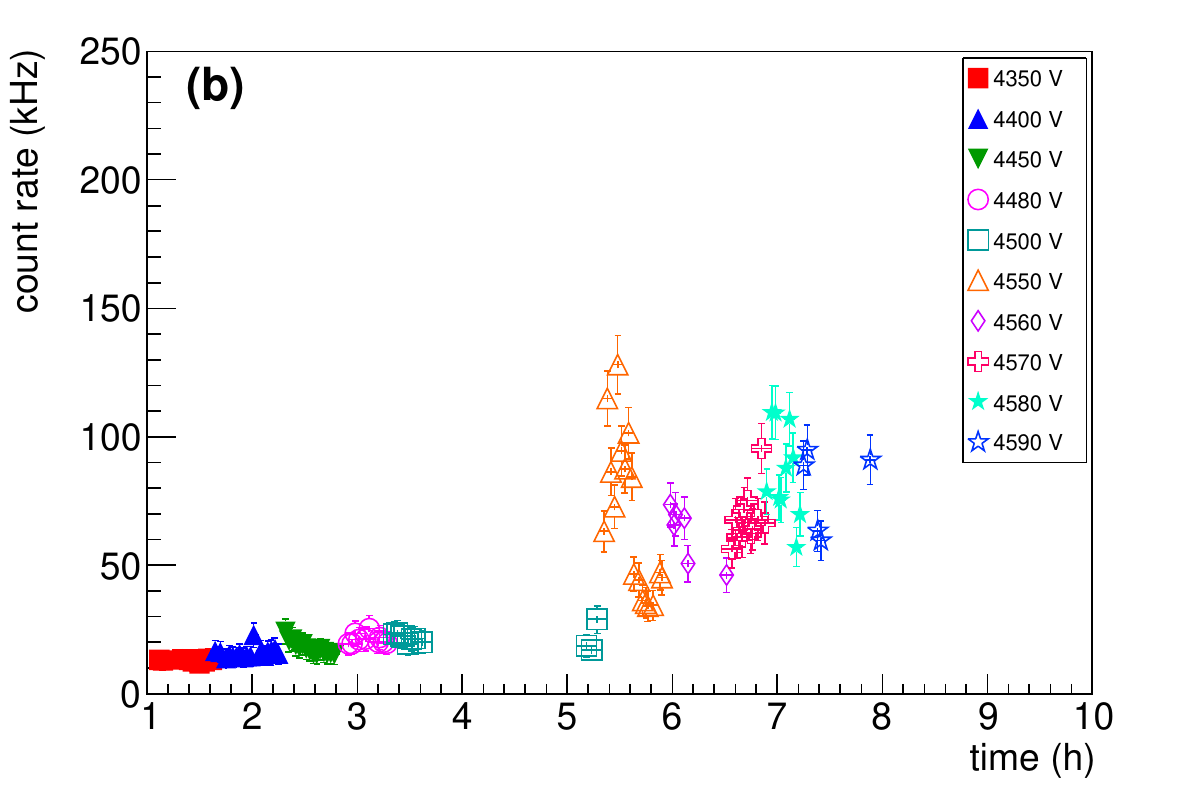}
	\label{fig:sub2}
	\end{subfigure}
	
	\vspace{0.5em}
	\begin{subfigure}[b]{0.495\textwidth}
	\includegraphics[width=\linewidth]{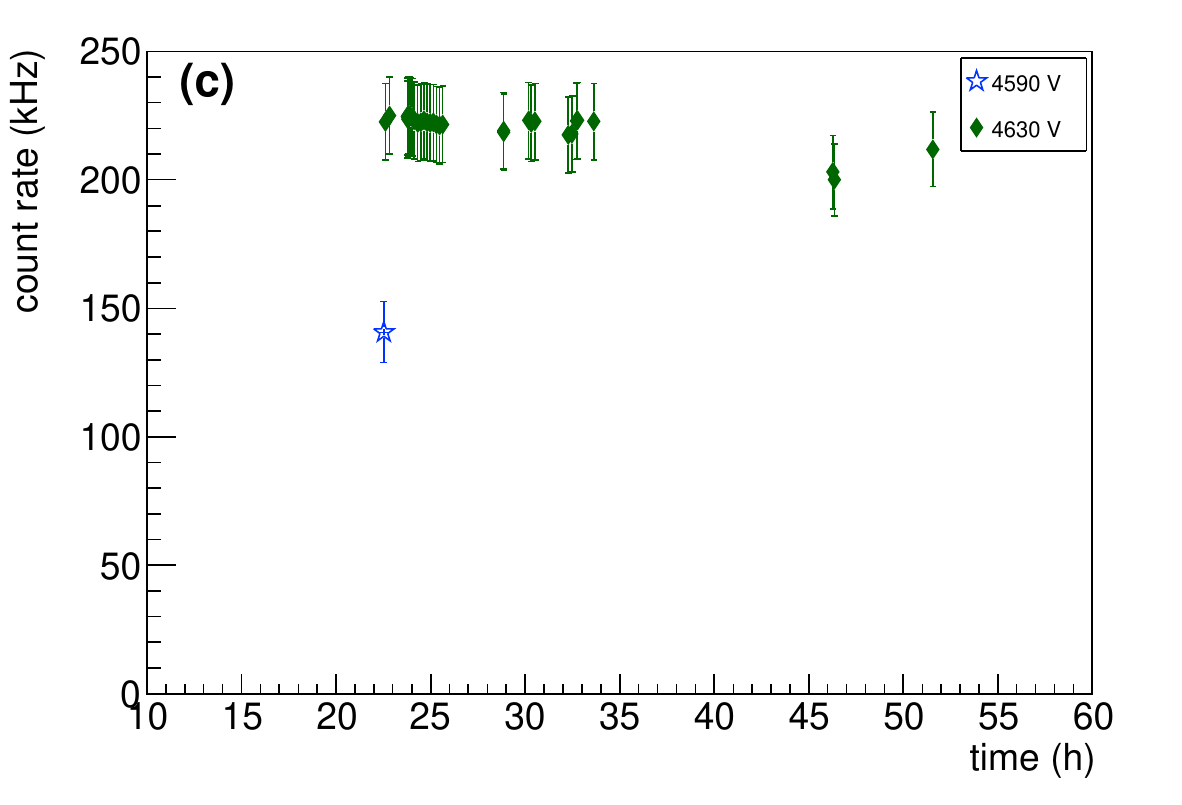}
	\label{fig:sub3}
	\end{subfigure}

	\caption{(Colour online) Variation of the count rate as a function of the time for three time zones, for 0-1 hour (6(a)), 1-10 hour (6(b)) and 10-55 hour (6(c))}\label{fig6}
\end{figure}
It is observed that during the first $\sim 8$ hours, the current and count rate increases with time. It is to be mentioned here that, the applied voltage is not kept constant as the current decreases with time after reaching a maximum, rather the applied voltage is increased to keep the current as well the individual $\Delta V$ across the GEM foils constant. It can be seen from Fig.~\ref{fig4} that even after 20 hours the current continues to decrease, however, the count rate remains nearly the same. To check this correlation the count rate is plotted as a function of current in Fig.~\ref{fig5} for all the voltage settings. Except some initial points at 4300 V when the detector is in conditioning phase the count rate increase with current. It is observed from  Fig.~\ref{fig5} that for the currents between $695\text{--}720~\mu\text{A}$ the count rate increases linearly but slowly from $10\text{--}20~\text{kHz}$ but after $720~\mu\text{A}$ the slope of this increase changes rapidly. For the highest voltage settings of 4630~V and at divider current $\sim717\text{--}722~\mu\text{A}$ maximum count rate  $\sim 220 \, \text{kHz}$ is recorded. It can be considered that in such settings the chamber is in its highest efficiency. To investigate the variation of the count rate with time, in Fig.~\ref{fig6} the count rate as a function of the time is plotted for three different time zones, for 0-1 hour (6(a)), 1-10 hour (6(b)) and 10-55 hour (6(c)) respectively. It can be seen from Fig. 6(a) that immediately after the start of operation the first 0.4 hour count rate is high $(\sim 200 \, \text{kHz})$. After that it decreases rapidly. Subsequently the voltage is increased and the divider current and count rate also increase. However, for each step of increase of voltage, even the current initially increases but it again starts decreasing. As a result to keep the bias current same the voltage is further increased. This conditioning phase continues for about 8 hours. At the end of 8 hours the count rate is measured to be $\sim 100 \, \text{kHz}$ at an applied voltage of 4590~V as shown in Fig. 6(b). From Fig. 6(c) it is seen that after $\sim20$ hours the count rate is measured to be $\sim 150 \, \text{kHz}$ at voltage 4590~V. For this point the current is found to be $\sim715~\mu\text{A}$. The voltage is further increased to 4630~V and at this voltage count rate is measured to be between $\sim 200 \, \text{kHz}$ to $\sim 220 \, \text{kHz}$. This voltage is kept constant for next 45 hours and the count rate is measured to be nearly unchanged, although the current was decreasing slowly as seen from Fig.~\ref{fig4}. At this high voltage level as the count rate remains constant at  $\sim 220 \, \text{kHz}$, it can be considered that the detector reaches its highest efficiency level which remains stable for the remaining period of operation.

\section{Summary and Discussion}
With a goal of the stability study, a SM triple GEM detector is characterised by Ar/CO$_2$ gas mixture in $70/30$ volume ratio and irradiating a $50~\text{mm}^2$ area of the chamber by $^{55}$Fe X-ray. Spectra are recorded every minute, along with the recording of temperature, pressure, and relative humidity using an in-house built data logger. The divider current is monitored manually from the HV power supply and the count rate is also measured. The results indicate that immediately after applying HV, the current increases rapidly, reaches a maximum value, and then begin to decrease. During this phase, the detector’s spectral response is found to be unstable, making it impossible to fit by a Gaussian function for gain and energy resolution calculations. The conditioning phase of the detector continues during the HV ramp until the voltage reaches $4630~\text{V}$ after about 23 hours. The gain is not stable during the first 10 hours, and energy resolution cannot be determined. However, after this period, the gain is calculated and showed an inverse correlation with the energy resolution. While the temperature and pressure fluctuations contribute to the variation in gain, the primary influencing factor of gain decrease is identified as the decrease in the bias current. In this particular study, the count rate of the chamber using a radioactive source, as a figure of a merit of efficiency, is checked in detail. It is observed that at the very beginning of operation at 4300~V the count rate is measured to be very high for the first 0.4 hour and after that it starts to decrease to a very low value. Subsequently, the count rate increases when the voltage is increased. The unstable behaviour of the count rate within the first 8 hours of operation is not only due to the initial conditioning phase but also a combined effect of charging-up and polarisation of the dielectric medium of the GEM foil. Further analysis over a longer period (20–50 hours) showed that the count rate stabilises with value $\sim 220 \, \text{kHz}$. The count rate with the radioactive source or the efficiency of the chamber is found to be low during the conditioning phase. At the highest voltage setting the count rate remains nearly constant even though at such high voltage the bias current slowly decreases. For a system where GEMs are used for tracking, the stability in efficiency is most important. Even though for a slight change in ambient temperature and pressure if the gain changes or for a decrease in the divider current if the gain decreases a bit but measured count rate remains constant, then such a detector will be useful for the tracking purpose in the high energy physics experiments. That is why further stability study for efficiency will be continued in the laboratory with strong radioactive source for longer period. The take home massage of this study is that instability in the count rate is observed along with instability in the gain, energy resolution during the conditioning phase. However, after 10 hours of operation, the chamber's efficiency (count rate) is found to be stable and does not depend significantly on the variation of the bias current or other ambient parameters even though the gain decreases. 

\section{Acknowledgements}
	
	The authors would like to thank the RD51 collaboration (presently DRD1 collaboration) for the support in building and initial testing of the chamber at CERN. The authors would also like to thank Mr. Subrata Das for helping in building the collimators used in this study. This work is partially supported by the research grant SR/MF/PS-01/2014-BI from DST, Govt. of India, and the research grant of the CBM-MuCh project from BI-IFCC, DST, Govt. of India. S. Mandal acknowledges his UGC-NET fellowship for the support. S. Biswas acknowledges the support of the DST-SERB Ramanujan Fellowship (D.O.No. SR/S2/RJN-02/2012).

\end{document}